\documentclass[a4paper]{llncs}

\newtheorem{madefinition}{Définition}

\newtheorem{monexp}{Exemple}

\usepackage{graphicx}
\usepackage{amssymb}

\usepackage{listings}

\usepackage[T1]{fontenc}
\usepackage{amsmath}

\begin{document}

\title{Vérification Formelle  des Processus Workflow Collaboratifs}

\titlerunning{Vérification Formelle  des Processus Workflow Collaboratifs}

\author{Zohra Sbaï\inst{1} et Kamel Barkaoui\inst{2}}

\authorrunning{Z. Sbaï and K. Barkaoui}

\institute{Universit\'{e} de Tunis El Manar,\\ Ecole Nationale d'Ing\'{e}nieurs de Tunis, \\
BP. 37 Le Belv\'{e}d\`{e}re, 1002 Tunis, Tunisia\\
\email{zohra.sbai@enit.rnu.tn
}\\
\and
Conservatoire National des Arts et M\'{e}tiers \\
Rue Saint Martin, 75141 Paris, France\\
\email{barkaoui@cnam.fr}
}

\toctitle{~~}
\tocauthor{~~}
\maketitle

\noindent\textit{RESUME}. Dans ce papier, nous proposons une méthode de vérification des processus workflow collaboratifs basée sur les techniques de model checking. En particulier, nous présentons une approche de vérification des propriétés de cohérence de ces processus en utilisant SPIN model checker. Nous traduisons d'abord la spécification des workflows adoptée (à savoir les WF-nets) vers Promela qui est le langage de description des modèles à vérifier par SPIN. Ensuite, nous exprimons les propriétés de cohérence en Logique Linéaire Temporelle (LTL) et utilisons SPIN pour tester si chaque propriété est satisfaite par la spécification Promela du WF-net en question. Enfin, nous exprimons les propriétés de k-cohérence pour les WF-nets modélisant plusieurs instances et de (k,R)-cohérence pour les processus workflow concurrents et qui possèdent des ressources partagées.\\

\noindent\textit{ABSTRACT}. In this paper, we present a method of verification of collaborative workflow processes based on model checking techniques. In particular, we propose to verify soundness properties of these processes using SPIN model checker. First we translate the adopted specification of workflows (i.e. the WF-net) to Promela which is the description language of models to be verified by SPIN. Then we express the soundness properties in Linear Temporal Logic (LTL) and use SPIN to test whether each property is satisfied by the Promela model of the WF-net in question. Finally, we express the properties of k-soundness for WF-nets modeling multiple instances and (k,R)-soundness for workflow processes with  multiple instances and sharing resources.\\

\noindent\textit{MOTS-CLES} : Vérification sur modèle, workflow collaboratif, SPIN, Logique Temporelle Linéaire, contraintes de ressources.\\


\newpage
\section{Introduction}

Un processus métier consiste en un nombre de tâches à assurer et l'ensemble des conditions qui déterminent leur ordre. Un processus métier qui met en jeu différentes entreprises partenaires s'appelle processus collaboratif. L'automatisation de ce processus métier, en tout ou en partie résulte en un workflow (WfMC, 1999). En effet, c'est une représentation spécifique pour laquelle les mécanismes de coordination entre activités, applications ou participants peuvent être gérés par un système de gestion de workflow (WfMS). Le nombre des WfMSs émergents est en croissance rapide, et par conséquent le besoin de mécanismes efficaces et d'outils de modélisation et d'analyse des processus workflow est crucial. Dans ce contexte, nous nous focalisons à la vérification formelle des processus workflow.

Une des plus puissantes méthodes formelles est le model checking ou la vérification sur modèle. Le principe de la vérification sur modèle est de générer tous les états possibles d'un programme et de vérifier si une propriété est vérifiée dans chaque état. En pratique, des algorithmes sophistiqués qui sont basés sur la théorie des automates et sur la logique sont nécessaires pour effectuer la vérification sur modèle. Cette vérification peut être réalisée automatiquement par un outil logiciel dit model checker.
Cet  outil vérifie si une certaine propriété est satisfaite en examinant d'une manière systématique tous les scénarios possibles du système. Il existe plusieurs model checkers puissants tels que NuSMV (Cimatti et al., 2002), BLAST (Henzinger et al., 2003) et SPIN (Holzmann, 2003) (Holzmann, 1997).
Ce dernier est un model checker développé par Gerard J. Holzmann pour vérifier les protocoles de communication. Il a été très utilisé dans les industries qui conçoivent des systèmes critiques. Les modèles à vérifier par SPIN sont écrits en Promela, un langage dans lequel est décrit le comportement du processus.

Dans cet article, nous proposons une méthode de modélisation et de vérification des processus workflow à l'aide de SPIN. Pour vérifier un système, nous avons besoin de décrire deux choses: l'ensemble des faits que nous voulions vérifier, et les aspects pertinents du système qui sont nécessaires pour vérifier ces faits. Ces aspects constituent le modèle à vérifier par SPIN, écrit en Promela et les faits représentent les propriétés à vérifier et qui doivent être spécifiés dans la Logique Temporelle Linéaire.

Tout d'abord, nous présentons une méthode pour la spécification du modèle de workflow dans le langage Promela. Ensuite, nous exprimons en LTL la propriété de cohérence des processus workflow qui exprime les exigences sur le comportement du système. Enfin, le model checker SPIN est exécuté pour vérifier si cette propriété est satisfaite. Dans le cas d'une réponse négative, SPIN fournit un contre-exemple montrant une exécution qui ne satisfait pas cette propriété.

La suite du papier est organisée comme suit. La section 2 présente des préliminaires sur les réseaux de Petri et la vérification sur modèle. La troisième section est consacrée au modèle Promela proposé des processus workflow. Dans la section 4 nous détaillons la vérification de la propriété de cohérence par SPIN. Nous étudions des aspects de cohérence pour les workflow complexes dans la section 5. Finalement, nous résumons notre approche en la comparant aux travaux existants et donnons des perspectives futures dans la section 6.

\section{Préliminaires}
Dans cette section, nous introduisons,  d'abord, la modélisation des processus workflow par les réseaux de Petri (RdP) (Barkaoui et al., 2007) ainsi que les définitions de base de ces derniers. Ensuite, nous présentons le principe du model checking.

\subsection{Modélisation des processus workflow}

Vu que les processus sont un facteur dominant dans la gestion des workflows, il est important d'utiliser et d'établir un cadre pour la modélisation et l'analyse de ces processus. Une variété de langages de modélisation de processus est disponible pour leur spécification. Ces langages peuvent être classés en deux types: ceux avec une représentation graphique (e.g. BPMN (OMG, 2008) et WF-nets (van der Aalst, 1997) et ceux avec une représentation textuelle (e.g. jpdl (JBoss, 2003), BPML (BPMI.org, 2002),
BPEL (Curbera et al., 2002) et XPDL (WfMC, 2002)).

Nous avons choisi d'adopter les WF-nets pour la modélisation des processus workflow. Les WF-nets forment une sous classe des RdPs dédiée à la gestion des workflow. En fait, les RdPs ont été utilisés pour leur sémantique formelle à caractère graphique, leur expressivité et la disponibilité des techniques et d'outils d'analyse.

Un réseau de Petri est défini par un quadruplet $N=(P,T,F,W)$ où \textit{P} et
\textit{T} sont deux ensembles non vides de places et de transitions respectivement, $P\cap T = \emptyset$, \\
$F\subseteq (P\times T) \cup (T\times P)$ est la relation flux et $W : (P \times T) \cup
(T\times P) \rightarrow \mathbb{N}$ est la fonction poids de
\textit{N} satisfaisant $W(x,y) = 0
\Leftrightarrow (x,y) \notin F$.
Si $W(u)=1$ $\forall u\in F$ alors \textit{N} est dit un réseau ordinaire et est denoté par $N = (P,T,F)$.

Pour tout $x\in P\cup T$, l'ensemble des entrées de \textit{x} est
$^{\bullet}x=\{y|(y,x)\in F\}$
et l'ensemble des sorties de \textit{x} est $x^{\bullet} =\{y|(x,y)\in F\}$.

Un marquage d'un RdP $N$ est une fonction $M:P\rightarrow \mathbb{N}$. Le marquage initial de $N$ est dénoté par $M_{0}$.
Une transition $t\in T$ est active (franchissable) dans un marquage $M$ (noté $M[t\rangle$) si et seulement si pour tout
$p \in$ $^{\bullet}t: M(p) \geq W(p,t).$
Si $M[t\rangle$, elle peut être franchie, et ceci résulte en un nouveau marquage $M'$ tel que :
$\forall p \in P:$ \\
$M'(p)=M(p)-W(p,t)+W(t,p)$. Ce franchissement est noté par $M[t\rangle M'$.

L'ensemble de tous les marquages accessibles de $M$ est noté $[M\rangle $.
Pour une place \textit{p} de \textit{P}, on note $M_{p}$ le marquage donné par $M_{p}(p)=1$ et
$M_{p}(p')=0$ $\forall p'\neq p$.
Les résaux de Petri sont représentés comme suit :
les places sont représentées par des cercles, les transitions par des rectangles, la relation flux est représentée par un arc entre \textit{x} et \textit{y} pour chaque $(x,y)$ dans la relation. Un marquage \textit{M} d'un réseau de Petri est représenté par l'ajout de $M(p)$ jetons noirs dans le cercle représentant la place \textit{p}.

Puisque les tâches d'un processus workflow doivent être exécutées dans un certain ordre, il est nécessaire d'introduire dans la définition des processus workflow des blocs de construction tels que ceux modélisant une séquence, des blocs conditionnels, parallèles et itératifs.
Il est donc clair qu'un réseau de Petri peut être utilisé pour spécifier le routage des processus. Les tâches sont représentées par les transitions et leurs dépendances causales par les places.
En fait, une place correspond à une condition qui peut être utilisée comme pré- et/ou post-condition pour les tâches.
Un réseau de Petri qui modélise un processus workflow est dit WorkFlow net (WF-net) (van der Aalst, 1997).

\begin{madefinition}

Un RdP $N=(P,T,F)$ est dit un workflow net (WF-net) ssi:
\begin{enumerate}
  \item $N$ possède une place source $i$ (i.e. $^{\bullet}i =\emptyset $) appelée place initiale.
  \item $N$ possède une place puit $f$ (i.e. $f^{\bullet} = \emptyset$) appelée place finale.
  \item pour chaque noeud $n\in P\cup T$, $\exists$ un chemin de \textit{i} vers \textit{n} et un chemin allant de \textit{n} à \textit{f}.
\end{enumerate}

\end{madefinition}

\subsection{Vérification sur modèle}

Model checking ou vérification sur modèle est une technique de verification qui explore exhaustivement tous les états possibles d'un système. Un model checker, l'outil logiciel qui assure la vérification sur modèle, examine tous les scénarios possibles systématiquement.

Nous proposons d'utiliser dans ce papier un des principaux model checkers qui est SPIN. Ce dernier est développé par Gerard J. Holzmann pour vérifier les protocoles de communication. Il a été très utilisé dans les industries qui réalisent des systèmes critiques. SPIN a été choisi à l'origine comme un acronyme pour Simple Promela INterpreter. En effet, les modèles devant être vérifiés par SPIN sont écrits en Promela. Ce dernier est un langage de modélisation en faisant des abstractions des systèmes distribués en supprimant les détails qui ne sont pas liés à l'interaction des processus. L'utilisation prévue de SPIN est de vérifier le comportement des processus. Ce comportement est modélisé en Promela et vérifié par SPIN.

La vérification se fait sur des propriétés exprimées en Logique Temporelle Linéaire (LTL).
LTL est construit en se servant d'un ensemble de variables propositionnelles $p_{1}, p_{2}, ... $, des connecteurs logiques usuels ($\neg$, $\vee$, $\wedge$, $\rightarrow $ et $\leftrightarrow$) et des opérateurs modaux temporels ($\lozenge$ :
éventuellement, $\square$ : toujours, $\circ$ : prochain et $\mathfrak{U}$ : jusqu'à)

La sémantique, le sens, d'une formule syntaxiquement correcte est défini en lui donnant une interprétation (une attribution de valeurs de vérité, T (true) ou F (faux)) à ses propositions atomiques et une extension de l'attribution à une interprétation de la formule entière selon des règles reliées aux opérateurs. Pour le calcul propositionnel celles-ci sont données par les tables de vérité familiers.

\section{Un modèle Promela des WF-nets}\label{section:wfnettopromela}

\begin{figure}[h!]
\centering
\includegraphics[scale=0.4]{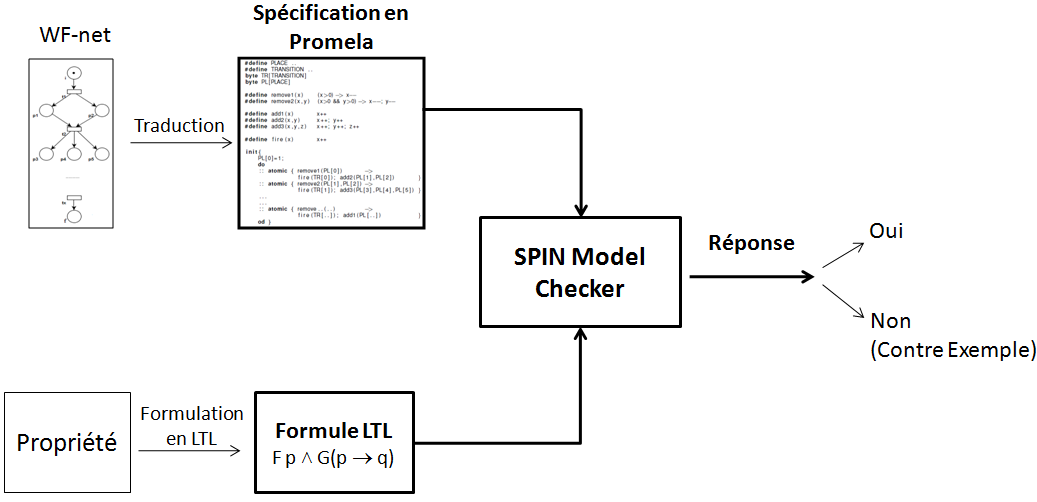}
\caption{Approche proposée de vérification des processus workflow}
\label{fig:demarche}
\end{figure}

Pour vérifier un système workflow, nous avons à décrire en LTL l'ensemble des propriétés devant être vérifiées. Mais, avant une telle description, il faut caractériser les aspects pertinents du système qui sont nécessaires pour vérifier ces propriétés. Ces aspects représentent le modèle de workflow et doivent être exprimés en Promela. La figure \ref{fig:demarche} représente les étapes de la méthode de vérification proposée.

Dans le reste de cette section, nous présentons notre approche de spécification d'un modèle de workflow en Promela et l'illustrons par un exemple.

Promela (PROcess MEta LAnguage) est un langage de spécification des systèmes concurrents tels que les protocoles de communication. Il autorise la création dynamique de processus. La communication entre ces processus peut être effectuée par le partage de variables globales ou en utilisant des canaux de communication. 
En Promela, il n'y a pas de différence entre les instructions et les conditions.
Une instruction ne peut être exécutée que si elle est réalisable, une condition ne peut être adoptée que si elle est vraie. Sinon, le processus est bloqué jusqu'à ce que la condition devienne vraie.

Le comportement d'un processus workflow peut être décrit par ses états et leurs changements.
Le marquage est initialisé à $M_{i}$ et il est changé en suivant cette règle de transition: une transition
$t$ est active si chaque place d'entrée $p$ de $t$ est marquée. Une transition active $t$ peut franchir ou pas,
et le tir de $t$ enlève un jeton de chaque $p\in$ $^{\bullet}t$ et ajoute un jeton à chaque place de $t^{\bullet}$.

Ce comportement dynamique peut être décrit en Promela dans le processus $init$.
En fait le mot clé $init$ est utilisé pour déclarer un processus qui sera actif à l'état initial du système.
Dans ce processus, nous proposons de décrire les exécutions des tâches par une boucle $do$ dans laquelle chaque ligne spécifie un tir d'une transition (i.e. exécution d'une seule tâche). 
La séquence d'instructions composées doit être exécutée comme une unité indivisible, non entrelacée avec d'autres processus, et par conséquent nous précédons cette séquence d'instructions par le mot clé $atomic$.

Nous proposons de présenter le modèle Promela d'un WF-net par une BNF (Backus
Naur Form) dans laquelle $<Process>$ est le symbole de départ :

\begin{center}
\small
\begin{tabular}{lll}
\hline

$<Process>$ & $::=$ & $init~~\{$\\
                & & $<Initial\_Marking>$\\
                & & $do$\\
                & & $<Firings>$\\
                & & $od$\\
                & & $\}$\\
$<Firings>$ & $::=$ & $<Firing>~<Firings>$ \\
$<Firing>$ & $::=$ & $::~~atomic~~ \{~~ $\\
                & & $<Enabled\_Tk>~~\rightarrow~~<Fire\_Tk>$\\
                & & $\}$\\
\hline

\end{tabular}
\end{center}

~~\\Notons que $<Enabled\_Tk>~\rightarrow~ <Fire\_Tk>$,
$<Enabled\_Tk>$ est une guarde exprimant que $<Fire\_Tk>$ ne va être exécutée que si $<Enabled\_Tk>$ est vrai.

D'après ce qui précède, il est possible de décrire un WF-net
en fonction du marquage des places (à travers une table d'entiers $PL$) et du nombre de tirs des transitions (table d'entiers $TR$).

Le comportement dynamique d'un processus workflow est traduit en Promela en décrémentant (resp. incrémentant) les cases associées aux places contenant des jetons à consommer (resp. celles dans lesquelles des jetons vont être produits). Ceci est assuré par
les $macros$ $fire$, $add1$, $add2$,
..,$addS$, $remove1$, $remove2$, .., et $removeK$ où $S$ est le nombre maximum des places d'entrée et $K$ est le nombre maximum de possibles places de sortie. En Promela, les définitions de macro sont utilisées comme un simple mécanisme pour remplacer celui d'appel de procédures.
Donc pour franchir une transition $t$ qui possède $I$ entrées et $J$ sorties, nous appelons ces macros avec les arguments appropriés.

\begin{enumerate}
  \item $removeI(p_{1}, p_{2}, .., p_{I})$ teste si ses places d'entrée sont marquées ($PL[indice\_de\_p_{j}]>0$, $1\leq j\leq I$) et si cette condition est vérifiée un jeton sera enlevé de chaque paramètre.
  \item $fire(t)$ trace le franchissement de $t$ par incrémentation de $TR[indice\_de\_t]$.
  \item et finalement $addJ(p_{1}, p_{2}, .., p_{J}$) ajoute un jeton à chaque place de sortie.
\end{enumerate}

En utilisant ces macros, la séquence atomique \\
$<Enabled\_Tk>~\rightarrow~ <Fire\_Tk>$ est réecrite comme suit :

\begin{center}
\small
\begin{tabular}{lll}
\hline

$removeQ(PL[I_{1}],PL[I_{2}],..,PL[I_{Q}])$ & $->$ & $fire(TR[index\_of\_Tk]);$\\
 & & $addM(PL[O_{1}],PL[O_{2}],..,PL[O_{M}])$\\

\hline
\end{tabular}
\end{center}

Dans cette spécification Promela du franchissement d'une transition $Tk$, nous avons noté par $Q$
le nombre des places de $^{\bullet}Tk$ et par $M$ le nombre des places de $Tk^{\bullet}$. Nous avons aussi noté par $I_{1},I_{2},..,I_{Q}$ les indices des places d'entrée dans $PL$ et par $O_{1},O_{2},..,O_{M}$ ceux des places de sortie.

\section{Vérification de la propriété de cohérence}

Une propriété est exprimée en SPIN par un automate fini appelé
"never claim" qui est exécuté en même temps avec l'automate fini qui
représente le programme Promela. Spécifier une propriété
directement comme un never claim est difficile ; en
contre partie une formule écrite en LTL sera
traduite par SPIN en un never claim, qui sera plus tard utilisé pour la
vérification.

SPIN fournit un algorithme de conversion des formules LTL en automates de B\"{u}chi.
never claim représente le modèle Promela de cet automate de B\"{u}chi qui correspond à la formule LTL en question.
Les never claims sont utilisés pour spécifier le comportement du système
qui ne doit jamais se produire.
Lors de la vérification, le model checker SPIN crée un processus représentant
le never claim déclaré. Le vérificateur exécute le processus du
never claim entre les exécutions de tous les autres processus et
reporte une erreur dans le cas où le processus associé au never claim se termine.

Donc, il nous suffit de spécifier nos propriétés en LTL pour qu'elles soient vérifiées par SPIN. Dans ce qui suit, nous nous concentrons à la propriété de cohérence  (van der Aalst, 1997) d'un workflow.

Cette propriété exige que pour n'importe quelle instance, elle va éventuellement  terminer (1: Terminaison) et au moment de la terminaison, il doit y avoir un jeton dans la place finale $f$ et toutes les autres places doivent être vides (2: Terminaison Propre).
De plus, il faut s'assurer de l'absence de transitions mortes, i.e. il doit être possible d'exécuter n'importe quelle tâche en
suivant un itinéraire spécifique sur le WF-net (3: Absence d'interblocage). Donc pour montrer la cohérence d'un processus workflow, il suffit de vérifier ces tois sous-propriétés.

Examinons chacune des sous-propriétés de cohérence à part afin de dégager leurs formules LTL.

\begin{itemize}
  \item \textbf{Terminaison}: la terminaison implique que "éventuellement l'état $M_{f}$ est atteint". La spécification de cette propriété en LTL peut donc être $\lozenge~term$ (lue éventuellement term) avec $term$ une proposition définie en Promela assurant que la place $f$ est marquée.
  \item \textbf{Terminaison Propre}: cette propriété peut intuitivement être interprétée comme suit: "si la place $f$ est marquée"(p1) alors "f contient exactement 1 jeton et toutes les autres places sont vides"(p2). Ceci peut être exprimée en LTL par la formule suivante:  $\square(term~\rightarrow ~prop)$ avec term une proposition assurant p1 et prop une proposition assurant p2.
  \item \textbf{Absence d'interblocage}: cette propriété exige que chaque transition est franchie au moins pour une exécution du workflow. Ceci revient à vérifier le franchissement de toutes les transitions du réseau fermeture $N^*$ au moins une fois. En LTL, on peut l'exprimer par la formule $\lozenge~live$ avec $live$ une proposition assurant le franchissement de toutes les transitions de $N^*$.
\end{itemize}

La propriété de cohérence est donc caractérisée en LTL comme suit :

\begin{madefinition}\label{def:sound}
Un WF-net $N$ est cohérent ssi les formules LTL suivantes sont satisfaites :
\begin{enumerate}
  \item $\lozenge~term$
  \item $\square(term~\rightarrow ~prop)$
  \item $\lozenge~live$ ~~~~ (pour $N^*$)
\end{enumerate}
\end{madefinition}

où $term$, $prop$ et $live$ sont des propositions définies en Promela comme suit :

\begin{center}
\fbox{
\begin{minipage}{0.8\columnwidth}
1) $\#define~term~(PL[|P|-1] >= 1)$\\
2) $\#define~prop~(\overset{i=|P|-2}{\underset{i=0}{\&\&}}  \; (PL[i]== 0)~\&\&~PL[|P|-1]== 1)$\\
3) $\#define~live~(\overset{j=|T|}{\underset{j=0}{\&\&}}  \; (TR[j]>=1))$   (For $N^*$)
\end{minipage}
}
\end{center}

Rappelons que $N^*$ est le réseau fermeture de $N$ obtenu en y ajoutant une transition $t^*$ et deux arcs reliant $f$ à $t^*$ et $t^*$ à $i$.

\begin{figure}[h!]
\centering
\includegraphics[scale=0.6]{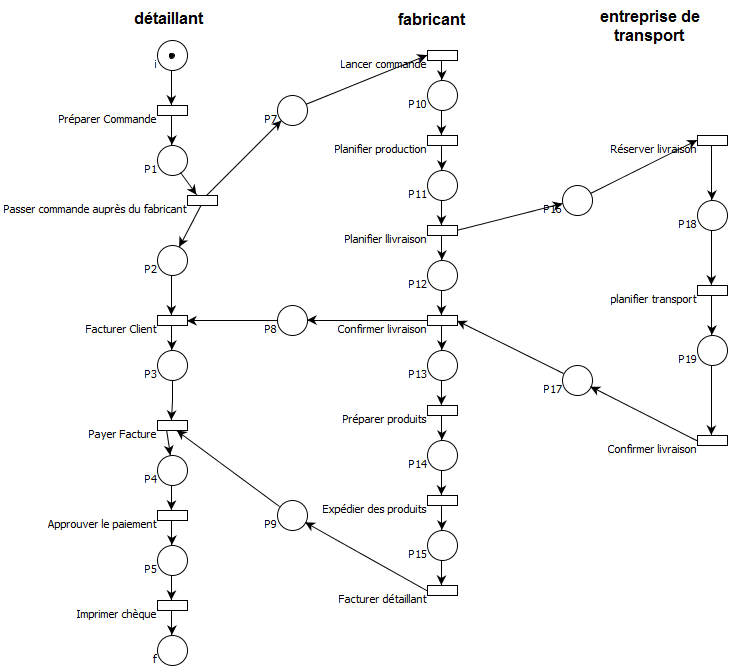}
\caption{Un WF-net résumant la gestion d'une chaîne logistique}
\label{fig:WFsupplychainmanuf}
\end{figure}

\begin{monexp} 
Nous considérons dans cet exemple le processus workflow collaboratif représentant la gestion d'une chaîne logistique modélisé par le WF-net de la figure ~\ref{fig:WFsupplychainmanuf}.
\end{monexp}

\begin{figure}[h!]
\centering
\includegraphics[scale=0.6]{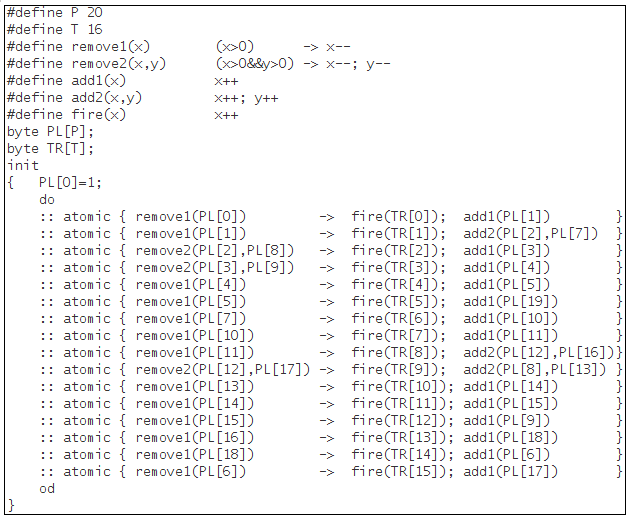}
\caption{Spécification Promela de la gestion d'une chaîne logistique}
\label{fig:PMLsupplychainmanuf}
\end{figure}

La spécification Promela de ce WF-net est donnée dans la figure ~\ref{fig:PMLsupplychainmanuf}.
Pour ce WF-net, les trois conditions de la définition \ref{def:sound} sont satisfaites et par conséquent il est cohérent.

\begin{figure}[h!]
\centering
\includegraphics[scale=0.6]{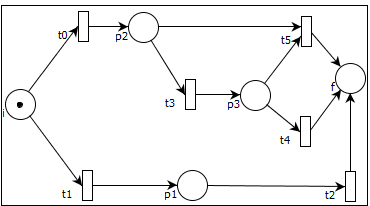}
\caption{Un exemple de WF-net non cohérent}
\label{fig:WF_netwsnots}
\end{figure}

\begin{figure}[h!]
\centering
\includegraphics[scale=0.36]{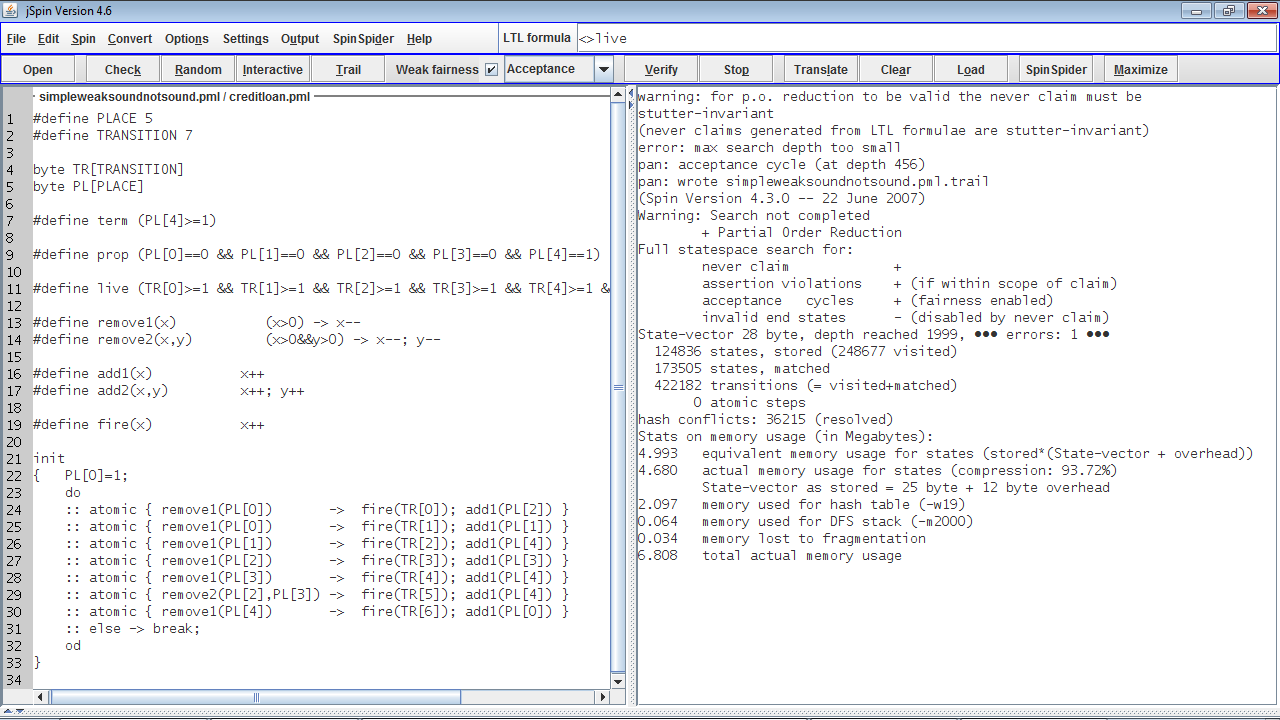}
\caption{Vérification de la non cohérence d'un WF-net}
\label{fig:wsnots2}
\end{figure}

\begin{monexp} Considérons maintenant le WF-net de la figure
~\ref{fig:WF_netwsnots}.
\end{monexp}
Ce WF-net n'est pas cohérent puisque la troisième condition de la définition \ref{def:sound} n'est pas satisfaite (figure \ref{fig:wsnots2}).

\begin{figure}[h!]
\centering
\includegraphics[scale=0.36]{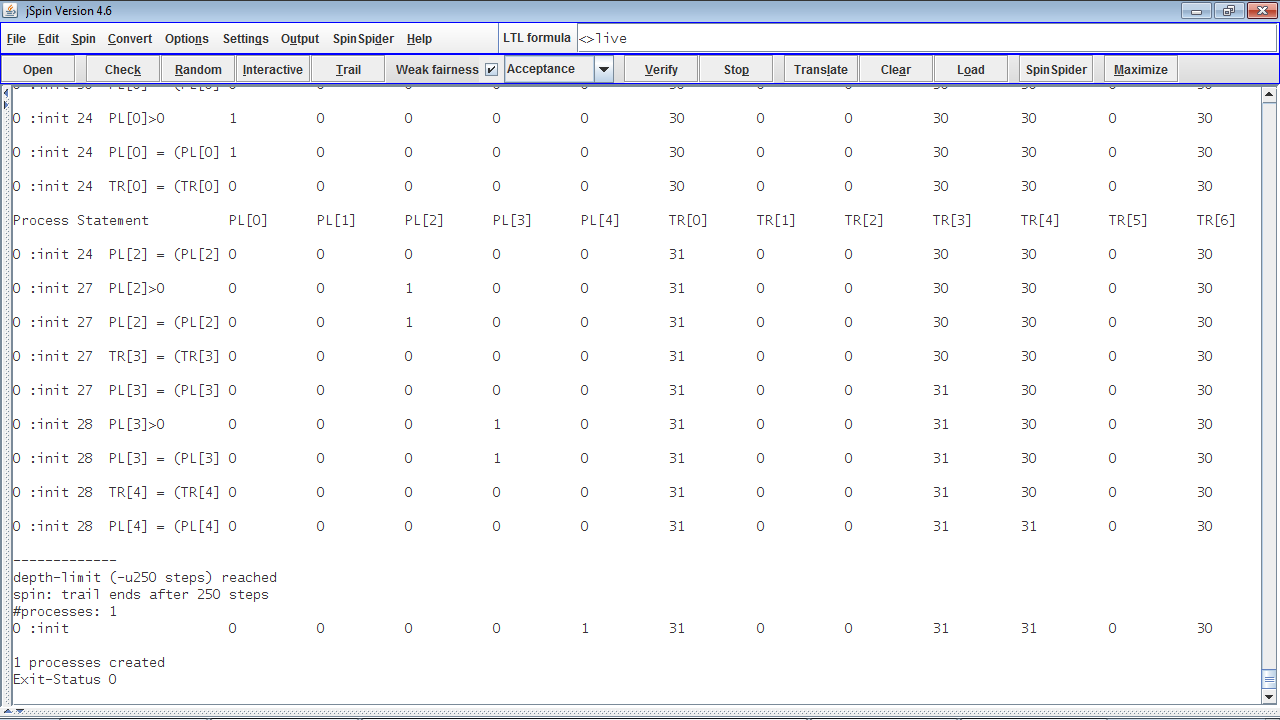}
\caption{Un contre exemple généré par SPIN} \label{fig:wsnots3}
\end{figure}

Un des points forts de la vérification sur modèle est le fait qu'on peut voir
un contre exemple dans le cas où une formule est violée.
Pour notre exemple, la figure ~\ref{fig:wsnots3} montre la première exécution dans laquelle l'état final ne guarantit pas la cohérence du WF-net.

Par ailleurs, il est trivial de noter que pour cet exmple le jeton présent dans la place $i$
peut toujours atteindre la place finale et au moment auquel un jeton est produit dans $f$, toutes les places autres que $f$ sont vides.
Par conséquent, les deux premières conditions de la cohérence sont satisfaites. On dit que le WF-net considéré satisfait la propriété de cohérence faible.

\section{Aspects de cohérence des processus workflow complexes}\label{sectsoundextension}

Les WF-nets étudiés dans la section précédente traitent une seule instance (cas) de la procédure modélisée (i.e. un jeton unique
existe au début à la place $i$). Dans cette section, nous considérons d'abord les processus workflow modélisant $k$ instances. Ensuite nous étudions les processus à ressources partagées.

    \subsection{k-cohérence des WF-nets modélisant des processus concurrents}

La propriété de k-cohérence est une
extension de la cohérence qui a été prouvé décidable dans
(Tiplea et al., 2005). La valeur $k$ ($k\in \mathbb{N}$) indique le nombre de jetons de
la place d'entrée $i$ à l'état initial et donc le nombre d'instances du
workflow qui sont prêtes à l'exécution. En conséquence, la k-cohérence
consiste en la terminaison propre des $k$ instances et à l'absence de transitions mortes.

Pour vérifier la k-cohérence par SPIN model checker, nous avons à modifier le tableau $PL$ par initialisation de son
premier élément à $k$.

\begin{madefinition}
Un WF-net $N$ est k-cohérent si et seulement si les formules LTL qui suivent sont satisfaites :
\begin{enumerate}
  \item $\lozenge~kterm$
  \item $\square(kterm~\rightarrow ~kprop)$
  \item $\lozenge ~live~$ ~~~~(pour $N^*$)
\end{enumerate}
\end{madefinition}

Avec kterm, kprop et live constituent des propositions définies comme suit :\\

\begin{center}
\fbox{
\begin{minipage}{0.8\columnwidth}
$\#define~kterm~(PL[|P|-1] >= k)$\\
$\#define~kprop~(\overset{i=|P|-2}{\underset{i=0}{\&\&}}  \; (PL[i]== 0)
~~\&\&~PL[|P|-1]== k)$\\
$\#define~live~(\overset{j=|T|}{\underset{j=0}{\&\&}}  \; (TR[j]>=1))$   (vivacité du réseau fermeture)
\end{minipage}
}
\end{center}

    \subsection{(k,R)-cohérence des WF-nets avec ressources partagées}

Vu l'intérêt de partage des ressources entre les différents partenaires dans un workflow collaboratif, nous proposons d'étendre les résultats trouvés dans la section précédente à couvrir la cohérence des processus
workflow avec ressources partagées.

Un processus workflow sous contraintes de ressources peut être représenté à l'aide d'un réseau de Petri appelé WFR-net (van Hee et al., 2005).
Dans ce RdP, les ressources doivent être préservées et les ressources disponibles
consistent en des jetons consommables par les instances du workflow.

$ki + R$ est le marquage initial du WFR-net où $R$ est le
marquage initial des places ressources. Si l'exécution de toutes les
instances du processus a proprement terminé, le
marquage final du WF-net souligné (ne considérant pas les
ressources) est atteint et ainsi les ressources sont libérées. Dans ce cas, le marquage final du WFR-net 
est $kf + R$.

La propriété de cohérence des WFR-nets qui modélisent $k$ instances et qui partagent des ressources disponibles (avec le marquage $R$) est appelée (k,R)-cohérence.
Pour vérifier la (k,R)-cohérence en utilisant SPIN, nous partons du modèle Promela des WF-nets que nous avons proposé
et nous le modifions par modélisation des
places ressources (une place par type de ressource) et initialisation de leur marquage. 

Notons que dans les WFR-nets, nous avons des arcs dont le poids est supérieur à 1. Donc il nous suffit d'étendre les macros $remove$ et $add$ afin qu'ils puissent être utilisés pour la consommation et la production d'un nombre quelconque de jetons à la fois dans une même place.
Nous proposons donc de les redéfinir comme suit :

\begin{itemize}
  \item $removeI(p_{1}, p_{2}, .., p_{I}, n_{1}, n_{2}, .., n_{I})$ détruit  $n_{j}$ jeton(s) de chaque place  d'entrée repérée par $p_{j}$ où $j$ varie de $1$ à $I$.
  \item $addJ(p_{1}, p_{2}, .., p_{J}, n_{1}, n_{2}, .., n_{J}$) crée  $n_{1}$ jeton(s) dans la place  $p_{1}$, $n_{2}$ jeton(s) dans la place  $p_{2}$, .., $n_{J}$ jeton(s) dans la place  $p_{J}$.
\end{itemize}

\section{Conclusion}

Nous avons proposé une méthode basée sur le model checker SPIN pour modéliser et analyser les processus workflow spécifiés par des WF-nets.
Nous avons, en premier lieu, détaillé comment traduire en Promela un processus métier donné. Ceci consiste en la définition des aspects servant comme matériel de vérification des processus, à savoir le tir des transitions et l'évolution d'états. La traduction proposée couvre les patrons workflow structurels complexes.

En second lieu, nous avons étudié différentes propriétés de cohérence des WF-nets et présenté leur formulation en LTL.
Ces propriétés ont couvert d'une part la cohérence des processus workflow représentant un cas et faisant abstraction des ressources partagées. D'autre part, nous avons considéré la cohérence des processus modélisant $k$ instances et partageant des ressources de différents types. Nous pouvons facilement vérifier d'autres propriétés telles que la propriété de safeness.

Les exemples étudiés ont montré l'importance de notre approche, et plus particulièrement dans le cas de violation d'une propriété. En effet, dans ce cas, un contre exemple est généré et peut être étudié en vue de correction. Par ailleurs, la méthode proposée peut être utilisée pour vérifier tout type de système complexe.

Dans la littérature peu sont les travaux concernent l'utilisation des techniques de model checking pour modéliser et analyser les
processus workflow. Dans (Conghua et al., 2006), les auteurs utilisent NuSMV et prouvent que CTL* peut être utilisée pour caractériser
la cohérence relaxée des WF-nets. Le travail dans (Yamaguchi et al., 2008) présente une méthode de vérification des processus workflow par SPIN et ce sur la base des résultats donnés dans (Holzmann, 2003).

Notre travail est semblable à celui de Yamaguchi et al. puisque nous utilisons SPIN model checker afin de vérifier les processus workflow. Cependant, notre contribution ne concerne pas seulement les WF-nets simples, mais aussi ceux modélisant des processus complexes tels que les WFR-nets.

Actuellement, nous travaillons sur l'extension de notre approche à l'analyse paramétrée des processus workflow à contraintes temporelles (Boucheneb et al., 2012).

\end{document}